\documentclass[aps,prb,twocolumn,groupedaddress]{revtex4}

\usepackage{amsfonts} 
\usepackage{amssymb}  
\usepackage{graphicx} 
\usepackage{graphics}
\usepackage[dvips,final]{epsfig}  
\usepackage{amsmath}  
\usepackage[latin1]{inputenc}
\usepackage{latexsym}
\usepackage[amssymb]{SIunits}

\begin{document}


\title{Two-stage dissipation in a superconducting microbridge:\\
Experiment and modeling}


\author{L. Del R{\'i}o, E. Altshuler}
\affiliation{Superconductivity Laboratory, IMRE-Physics Faculty,
University of Havana, 10400 Havana, Cuba}
\author{S. Niratisairak, O. Haugen and T. H. Johansen*}
\affiliation{Department of Physics, University of Oslo, 0316 Oslo, Norway}
\affiliation{*Institute for Superconducting and Electronic Materials, University of Wollongong, Australia}
\author{B. A. Davidson}
\affiliation{INFM-TASC Area Science Park, Basovizza, Italy}
\author{G. Testa, E. Sarnelli}
\affiliation{Cybernetic Institute of the CNR, Via Campi Flegrei 34,
80078, Pozzuoli (NA), Italy}



\date{\today}

\begin{abstract}
Using fluorescent microthermal imaging we have investigated the origin of ``two-step'' behavior in {\it I-V} curves for a current-carrying YBa$_2$Cu$_3$O$_x$ superconducting bridge. High resolution temperature maps reveal that as the applied current increases the first step in the voltage corresponds to local dissipation (hot spot), whereas the second step is associated with onset of global dissipation throughout the entire bridge. A quantitative explanation of the experimental results is provided by a simple model for an inhomogeneous superconductor, assuming that the hot spot nucleates at a location with slightly depressed superconducting properties.\\[0 cm]
\end{abstract}

\pacs{???}

\maketitle

	\section{Introduction}

Jumps and other discontinuities in the current-voltage ({\it I-V}) characteristics of superconductors are commonly regarded as a fingerprint of failure. Their relevance for the behavior of superconducting magnets, storage coils and power transmission lines makes them a widely studied subject in the literature, see e.g., Ref.~[\onlinecite{Huebener79}]. While some kinds of discontinuities in {\it I-V} curves are claimed to be indications of self organization in the dynamics of moving vortices in the presence of microscopic disorder~\cite{Nori98,Bassler01,RMP04}, there is wide consensus that ``catastrophic" jumps are associated with resistive heating of macroscopic parts of the sample~\cite{Gurevich87,Jardim96}. In the latter case, finite regions of the superconductor turning normal as the applied current increases have been assumed responsible for local overheating in order to explain the jump characteristics~\cite{Gurevich87}. While this ``hot spot" scenario constitutes a sound physical picture, it is still highly desirable to actually visualize these regions in a current-carrying superconductor to obtain more direct evidence and better understanding of the relation between {\it specific} discontinuities during real {\it I-V} measurements and the actual origin and extent of the regions involved in the jumps.

In this work we report results from combined 4-probe transport measurements and visualization using the method of fluorescent thermal imaging (FTI)~\cite{Oyvind07,Oyvind08}, carried out on a thin film bridge. The experimental results are analyzed by comparing the combined observed behavior with a simple theoretical model for a bridge with a small region of depressed superconducting properties. For the two jumps found in the {\it I-V} curve, it is shown that the first corresponds to the appearance of a localized hot spot in the defected region, and the second to the onset of an overall dissipation throughout the bridge. The model, having only two adjustable parameters, gives an excellent quantitative description of the behavior.

	\section{Experimental results}

The sample was a $d = 0.3$~$\mu$m thick c-axis oriented film of YBa$_2$Cu$_3$O$_x$ (YBCO) with a superconducting transition temperature of $T_{\textrm{c}} = 91$~K. The film was photo-lithographically patterned into a $w= 5$~$\mu$m wide and $l = 500$~$\mu$m long strip, which at both ends extends into large areas coated with gold for electrical connection. Space-resolved observation of the dissipation in the bridge was obtained using FTI, where a 1~$\mu$m thick film of poly(\-methyl \-methacrylate) (PMMA) mixed with the fluorescent dye europium tris\-[3-(trifluoro\-methyl\-hydroxy\-methylene)\--(+)-\-camphorate]\- (EuTFC) deposited by spin coating on the sample, was used as sensor.
\begin{figure}[h!!]
  \includegraphics[height=7.5cm, width=7.0cm]{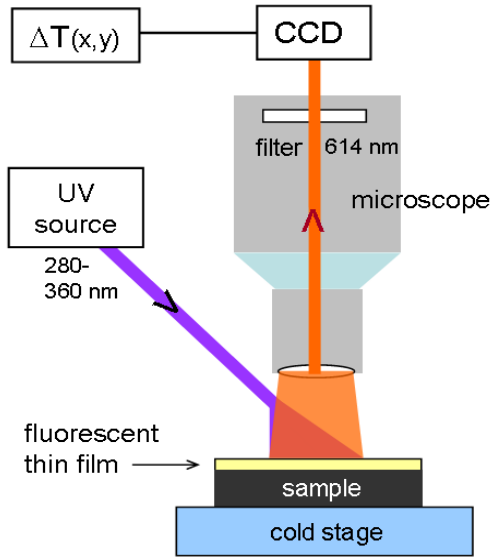}
  \caption{Experimental setup for micro-thermal imaging where a fluorescent polymer film serves as temperature sensor over the sample area.}\label{setup}\vspace{-0.2cm}
\end{figure}

Our FTI setup consists in a standard Leica DMR microscope and a continuous flow cryostat, where the sample is mounted on the cold finger below an optical window (Suprasil) allowing the incoming UV-light to excite the fluorescent film, see Fig.~\ref{setup}. The UV-source is a 200~W mercury-xenon lamp, a UV transmission filter and an optical light-guide. The light emitted from the sensor film at 614~nm has a strongly temperature dependent intensity, which through the microscope equipped with a 10~nm bandpass filter creates a direct map of the temperature distribution over the sample area. A more detailed description of the setup can be found elsewhere.~\cite{Oyvind07} The present experiments were carried out at $T_{0} = 84$~K, and the thermal images had a temperature resolution better than 100~mK.
\begin{figure}[b!!]
  \includegraphics[width=0.5\textwidth]{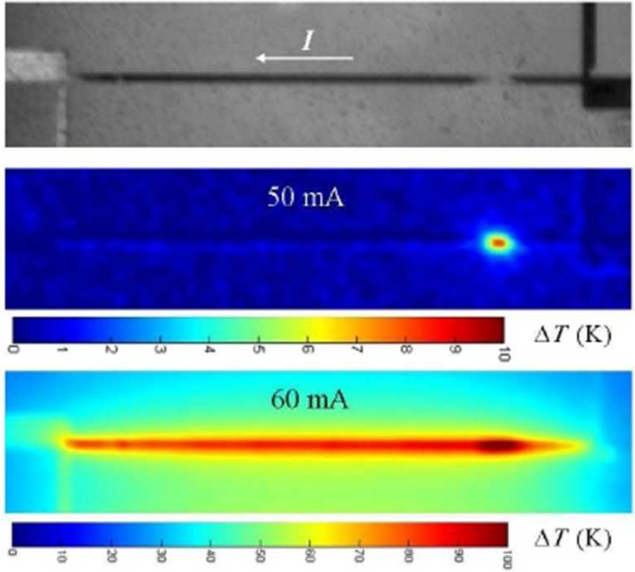}
  \caption{From top to bottom is shown an optical image of the bridge(horizontal strip) and two thermal images for applied currents of $50$~mA and $60$~mA respectively. Both thermal images have a color-coded scalebar where $\Delta T = 0$ corresponds to 84~K, the ambient temperature. The bridge is 0.5~mm long.}\label{D_T_Exp}\vspace{-0.2cm}
\end{figure}

\begin{figure}[h!!]
  \includegraphics[width=8.0cm]{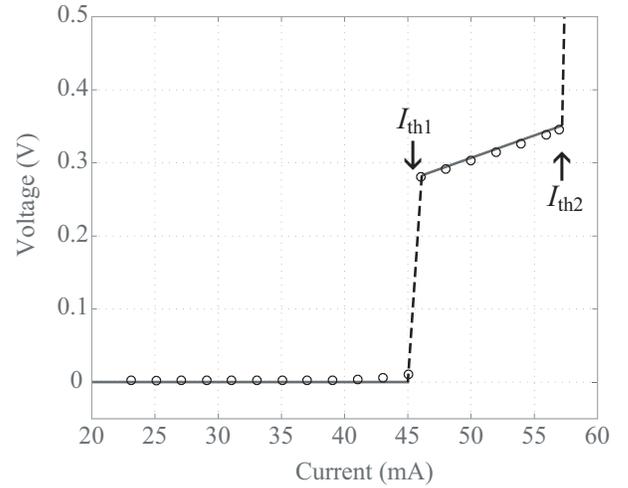}
  \caption{Plot of experimental current-voltage({\it I-V}) characteristic of the bridge shows voltage jumps appearing the threshold currents $I_{\textrm{th1}}$ and $I_{\textrm{th2}}$. The voltage respond for $I = 60~\textrm{mA}$ is higher than 20~V, excluded in this plot. The fitted lines are calculated from Eq.~\eqref{I-V-up}.}\label{I-V-Exp}\vspace{-0.2cm}
\end{figure}

A stabilized dc-current source was used to apply the transport current in the bridge. The 4-probe {\it I-V} measurements, and the recording of thermal images were performed simultaneously. Shown in Fig.~\ref{D_T_Exp}, middle and lower panels, are thermal images of the sample carrying a current of 50~mA and 60~mA, respectively. For reference, the upper panel shows an optical image of the bridge. Thermal images were recorded as the current was slowly ramped up in much smaller steps.

The first evidence of heating was the hot spot seen in the middle panel. At $I =50$~mA this locally heated region was found to be temporarily stable both in size and position, and the maximum temperature relative to the background was $\Delta T = 9$~K, which implies the temperature in the spot center was just above $T_{\textrm{c}}$. Note that any heat leakage into the bridge from external areas, such as the contact pads, was not observed. Increasing the current did not cause any immediate change in the thermal distribution. At $I =60$~mA a second dramatic event took place. Now the entire bridge became normal, and with an average temperature elevation of $\Delta T = 80$~K. The center of the earlier hot spot continues to have the highest temperature, now with $\Delta T = 100$~K.

This two-stage process is reflected also in the {\it I-V} curve shown in Fig.~\ref{I-V-Exp}. A sudden increase in the voltage by $\Delta V_1 = 280$~mV at $I_{\rm th1} =45$~mA obviously corresponds to the formation of the hot spot seen in the middle panel of Fig.~\ref{D_T_Exp}. The resistance of the hot spot is found to be $R_{\textrm{h}} = \Delta V_1 / I_{\rm th1} = 6.2~\Omega $. To a first approximation, the size $\Delta x$ of the region heated above $T_{\textrm{c}}$ can be estimated using that
$$
\Delta x/l =  R_{\textrm{h}}/R_{\textrm{n}}\ ,
$$
where $R_{\textrm{n}}$ is the normal state resistance of the bridge. From $R_{\textrm{n}}/l = \rho_{\textrm{n}}/wd$, and a normal state resistivity for YBCO~\cite{Xiao} of $\rho_{\textrm{n}} = 10^{-6}$~$\Omega$m, one obtains $\Delta x=9.3$~$\mu$m, i.e., twice the width of the bridge, which is fully consistent with the hot spot image in Fig.~\ref{D_T_Exp}, middle panel.

As the current is increased further, the voltage grows linearly and with a slope making the {\it I-V} curve extrapolate through origin. This constant resistance suggests a stable hot spot, as indeed was seen by the FTI. Then at $I_{\rm th2} = 57$~mA the voltage displays a second jump, which according to the thermal image at $I =60$~mA, brings the whole bridge well into the normal state.

	\section{Modeling and discussion}

To discuss the results in more detail a model was developed aiming to reproduce the reported two-stage dissipation process, as well as its full spatial and temporal evolution. We consider the long thin superconducting strip as one-dimensional, and aligned with the $x$-axis. It is assumed that once a part of the strip becomes normal, that part will be a source for Joule heating. The heat then propagates in two different ways, {\it(i)} by thermal conduction along the strip, and {\it(ii)} by escaping into the environment according to the Newton law of cooling with an effective heat transfer coefficient, $\alpha$. In the strip the temperature $T=T(x,t)$ then satisfies the following equation,
\begin{equation}
   c \frac{\partial T}{\partial t} =  k\frac{\partial^2 T}{\partial x^2} -  \frac{\alpha}{d} (T-T_{0}) + \frac{I^2\rho_{\textrm{n}}(x)}{w^2d^2}\Theta[T-T_{\textrm{c}}(x)]\, ,\label{heateq}
\end{equation}
with boundary conditions $T(0,t) = T(l,t)= T_{0}$. Here $c$ is the specific heat per unit volume and $k$ is the thermal conductivity. The unit step function, $\Theta$, adds the Joule heating term only for $x$ where $T$ is above the local transition temperature $T_{\textrm{c}}(x)$.

\begin{figure}[h!!]
  \includegraphics[width=8.5cm]{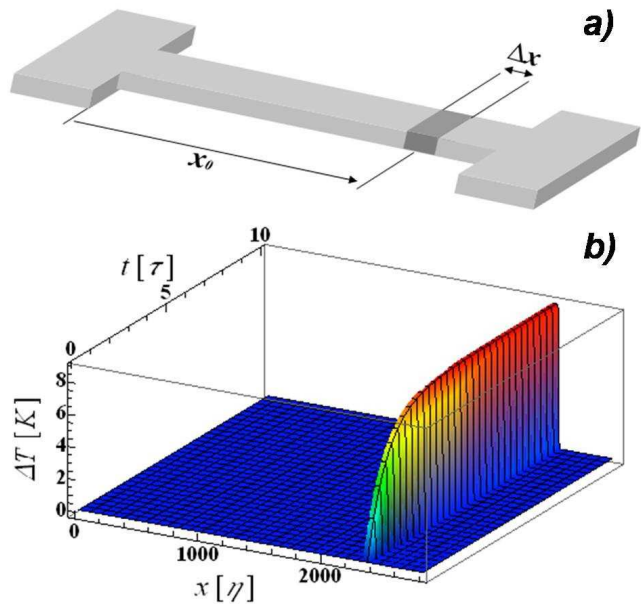}
  \caption{a) Sketch of the superconducting bridge where the position, $x_{0}$, and the length, $\Delta x$, of the ``weak" region are shown. b) Graph of the temperature distribution along the bridge and its dynamical behavior after applying a current of $I = 50~\textrm{mA}$, which leads to the formation of a hot spot. Units on the axises are $\eta=\sqrt{kd/\alpha}$ and $\tau=cd/\alpha$, the characteristic length and time, respectively.}\label{D_T_Teo}\vspace{-0.2cm}
\end{figure}

By including explicit coordinate dependences of $T_{\textrm{c}}$ and $\rho_{\textrm{n}}$, this model allows for nonuniformity. Indeed, comparing the direct image with the FTI images in Fig.~\ref{D_T_Exp}, a correlation is evident between the hot spot position and the region of optical nonuniformity (of unknown origin, but clearly visible). Actually, also a previous study~\cite{APL97} of a YBCO bridge where low-temperature scanning electron microscopy (LTSEM) was used to make a detailed map of the critical temperature, showed when compared with magneto-optical images of the bridge while passing a supercritical current, that the permanent damage directly correlated with the regions of depressed transition temperature. Motivated by this, we will assume the following coordinate dependencies,
\begin{equation}
 T_{\textrm{c}}(x) /T_{\textrm{c}} = 1- \frac{I/I_{0}}{1-\nu\Theta[(\Delta x/2)^2-(x-x_0)^2]} \, \label{Tc_I}
\end{equation}
and
\begin{equation}
 \rho_{\textrm{n}}(x) = \frac{\rho_{\textrm{n}}}{1-\nu\Theta[(\Delta x/2)^2-(x-x_0)^2]} \, ,\label{rho_x}
\end{equation}
where $\nu$ is a numerical factor, for simplicity set equal in both equations. The region of depressed properties is of size $\Delta x$ and centered at $x=x_0$, see Fig.~\ref{D_T_Teo} The relation also includes that the critical temperature gradually drops for large applied currents, where the parameter $I_{0}$ sets the scale.

Let us first use the present model together with the experimental results to determine the model parameters $\nu$ and $I_{0}$. As the applied current increases, dissipation will start in the ``weak" region once the current reaches the first threshold value, $I_{\rm th1}$. The onset happens as $T_{\textrm{c}}(x)$ in this region drops to $T_{0}$, and from Eq.~\eqref{Tc_I} it follows that,
\begin{equation}
 I_{\rm th1}=(1- {\nu})I_{0} (1 - T_{0}/T_{\textrm{c}})\, .\label{Ith1}
\end{equation}
When the current is increased further, also $T_{\textrm{c}}(x)$ for the remaining part of the bridge eventually falls below $T_{0}$, and the whole strip starts to dissipate. This second threshold, $I_{\rm th2}$, is given by
\begin{equation}
 I_{\rm th2} = I_{0}(1-T_{0}/T_{\textrm{c}})\, .\label{Ith2}
\end{equation}
From the experimental values $ I_{\rm th1}=45$~mA and $I_{\rm th2} = 57$~mA, we obtain $\nu = 0.21$ and $I_{0}=0.74$~A.

The form of Eq.~\eqref{heateq} defines a characteristic length $\eta = \sqrt{kd/\alpha}$ for spatial variations in the temperature profile. Provided that $\Delta x \gg \eta$ and the ``weak" region is dissipating, the temperature is effectively uniform within each homogeneous part of the strip. In particular, when passing an intermediate current $I_{\rm th1}<I<I_{\rm th2}$, it follows that the steady-state temperature rise of the hot spot equals,
\begin{equation}
 \Delta T = \frac{I^2\rho_{\textrm{n}}}{w^2 d (1-\nu) \alpha}\, .\label{DeltaT}
\end{equation}
Using for YBCO films, $k=5$~W/m\,K \!\cite{Marshall, Hagen} together with $\Delta T = 9~\textrm{K}$ revealed by the thermal image for $I$= 50~mA in Fig.~\ref{D_T_Exp}, the final model parameter is found to be $\alpha =4.7\cdot10^{7}$~W/m$^2$\,K. This value for the heat transfer coefficient is in line with values reported in the literature for similar samples~\cite{Nahum, Zeuner}. As a check for consistency, note that this gives $\eta = 0.18$~$\mu$m, which indeed satisfies $\eta \ll \Delta x$.

From this model it follows that the {\it I-V} characteristic of the bridge can be expressed as,
\begin{equation}\label{I-V-up}
  V(I)\!=\!\frac{I\rho_{\textrm{n}}}{wh}\!\left [\!\frac{\Delta x}{1-{\nu}}\Theta(I\!-\!I_{\rm th1})+
  (l\!-\!\Delta x)\Theta(I\!-\!I_{\rm th2})\!\right ],
\end{equation}
Here only the hot spot size is now adjustable, and best fit to the experimental data in Fig.~\ref{I-V-Exp} is obtained for $\Delta x = 7.3$~$\mu$m. The fitting result is included in the figure as straight lines. Direct inspection of the first and the second image in Fig.~\ref{D_T_Exp}, shows that $\Delta x = (8 ± 2)$~$\mu$m, in excellent agreement with the value obtained within the model.

With all model parameters found, we have used Eq.~\eqref{heateq} to determine the transient behavior of the hot spot by solving the equation numerically with $T(x,0)=T_{0}$ as initial condition. Real dimensions were used for the size of the bridge, where the position of the ``weak" region corresponds to $x_0=5l/6$, see Fig.~\ref{D_T_Teo}(a). The position along the bridge is measured in units of $\eta$, whereas the time unit is $\tau=cd/\alpha$. With a specific heat of $c = 1.2\cdot10^6$~J/m$^3$\,K for YBCO~\cite{Heatinger}, we have $\tau=7.6$~ns. Shown in Fig.~\ref{D_T_Teo}(b) is the resulting time evolutions of $\Delta T(x,t)\equiv T(x,t)-T_{0}$ along the bridge after a constant current of $50$~mA was turned on at $t=0$. Whereas the transient itself is much too fast to follow experimentally with the FTI method, we find a very good agreement between the model results and the experimental steady-state temperature profiles.

\section{Conclusions}

We have reported {\it direct} observation suggesting that two-step {\it I-V} curves in non-homogeneous superconductors can be caused by the sudden appearance of a ``hot spot" in a region with depressed superconductivity, followed by a stage in which the whole bridge passes to the normal state. We have constructed a simple model of the sample consisting in a superconducting bridge with a finite region of depressed critical temperature and increased normal state resistivity. The model reproduces quantitatively the two-step experimental {\it I-V} curve, as well as the corresponding thermal maps. Our combined experimental and theoretical results also suggest that ``multi-step", subsequent current {\it I-V} curves can be explained by the appearance of several ``hot spots" in regions of different superconducting properties connected in series into a superconducting sample---as commonly happens, for example, in a YBCO crystal with twinning defects, see e.g. Ref.~[\onlinecite{Jardim96}].

\section*{Acknowledgments}

The authors acknowledge the financial support from the Norwegian Research Council, and useful discussions with A. J. Batista-Leyva, O. Sotolongo-Costa and R. Jardim. The Max Planck Institute for the History of Science and the International Center for Theoretical Physics (ICTP) provided access to relevant scientific journals, and M. Arronte provided papers not available online.

\bibliographystyle{apsrev}

\end{document}